\documentclass{article}
\usepackage{mrs2005,epsfig}
\begin{document} 

\title{The Nuclear Accretion History of Red Galaxies} 
 
\author{K. Brand and the NDWFS and XBo\"otes teams}
\affil{National Optical Astronomy Observatory} 
 
\begin{abstract} 

We use an X-ray stacking technique to investigate the evolution of the mean X-ray luminosity of $\sim$21,000 red galaxies at $0.3\leq z < 0.9$ as a function of their stellar mass and redshift. The red galaxies are selected from the 9.3 deg$^2$ Bo\"otes field of the NOAO Deep Wide-Field Survey (NDWFS). The mean X-ray luminosity is an order of magnitude larger than we would expect from stellar sources alone and therefore must be primarily due to AGN emission. The X-ray luminosities ($L_x\approx 10^{41}{\rm ergs~s^{-1}}$) imply that these must be sources with relatively low accretion rates and/or accretion efficiencies onto their central super-massive black hole (SMBH). The mean X-ray luminosity increases significantly as a function of optical luminosity and stellar mass, implying that more massive galaxies have higher nuclear accretion rates than lower mass galaxies. We calculate that the mean X-ray luminosity evolves as $(1+z)^{3.3\pm1.5}$. This suggests a global decline in the mean AGN activity of normal early-type galaxies from $z\sim 1$ to the present. If accreting at typical AGN efficiencies, SMBHs within red galaxies accreted an insignificant proportion of their mass in this time. 

\end{abstract} 
 
\section{Introduction} 

The recent discovery that the mass of the central super-massive black hole (SMBH) within a galaxy is strongly correlated with the velocity dispersion and mass of its galactic bulge (Magorrian et al.~1998; Gebhardt et al.~2000; Ferrarese \& Merritt~2000) implies an intimate relationship between the build-up of galaxy bulges and the growth of their black holes by accretion processes. This suggests that the evolution of red (bulge dominated) galaxies may be linked to the evolution of their SMBH. Thus, determining the AGN activity in red galaxies as a function of redshift may help us understand how the galaxies themselves are evolving. 

Although deep X-ray exposures are typically needed to detect the individual X-ray flux from red galaxies out to $z\sim$1, by stacking the X-ray emission of large samples of galaxies with only shallow X-ray data, we can determine the X-ray characteristics on the ``mean object'' down to very faint limits. In this work, we use an X-ray stacking technique on $\sim$21,000 red galaxies selected from the NDWFS Bo\"otes field. This work is an extension of the work presented in Brand et al.~(2005) which was based on the Brown et al.~(2003) red galaxy sample in a 1.4 deg$^2$ sub-region of the NDWFS Bo\"otes field. A much larger red galaxy sample selected from the entire 9.3 deg$^2$ Bo\"otes field (Brown et al. in prep.) enables us to determine better constraints on their accretion history and to explore the relationship between the X-ray luminosity and other galaxy properties.
  
In this proceedings, we will determine whether the nuclear accretion rate is declining as a function of time in step with the decline in the SFR between $z$=1 and the present, and whether the SMBHs within red galaxies are accreting a significant proportion of their mass in this time. 
 
\section{Data} 

For this work, we make use of the optical data from the NOAO Deep Wide-Field Survey (NDWFS; Jannuzi \& Dey 1999) and the X-ray data from the XBo\"otes survey (Murray et al.~2005; Kenter~et~al.~2005; Brand~et~al.~2006) in the 9.3 deg$^2$ Bo\"otes region. The NDWFS is an optical and near-IR imaging survey to depths of $R\simeq 25.5$ and $K\simeq 19$. The XBo\"otes survey was observed by the Advanced CCD Imaging Spectrometer (ACIS) on the $Chandra$ X-ray Observatory and comprises 127 pointings of 5-ks each.

\begin{figure}
\begin{center}
\setlength{\unitlength}{1mm}
\begin{picture}(150,75)
\put(20,-15){\includegraphics{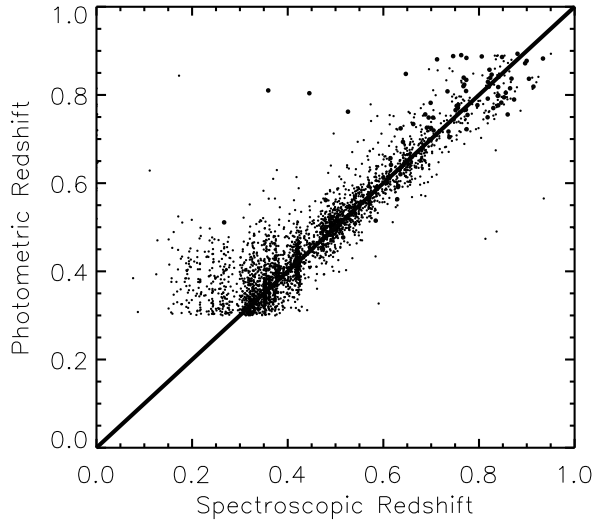}}
\end{picture}
\end{center}
{\caption[junk]{\label{fig:photoz} Photometric redshifts versus spectroscopic redshifts for all 3056 red galaxies with spectroscopic redshifts from AGES (small circles) and Keck (large circles) and photometric redshifts between 0.3 and 0.9.}}
\end{figure}

To select the red galaxy sample, we consider $B_W,R, $and $I$ imaging over almost the entire 9.3 deg$^2$ of the NDWFS Bo\"otes field (the field covered by our X-ray imaging). We determined photometric redshifts using the ANNz artificial neural networks code of Collister \& Lahav (2004), calibrated with a training set of 20,000 $I<20$ spectroscopic redshifts from the AGN and Galaxy Evolution Survey (AGES; Kochanek et al. in prep.) and several hundred $R<24.5$ Keck spectroscopic redshifts. We used the color-magnitude relation of galaxies to extrapolate our training set to faint magnitudes. Galaxy parameter such as absolute magnitudes, stellar masses, star formation rates (SFR), and tau values ($\tau$; the e-folding decay time of the SFR where the peak SFR is at high $z$) were determined by fits of PEGASE2 population synthesis models (Fioc \& Rocca-Volmerange 1997) to the NDWFS photometry. Our red galaxy sample comprises 21,388 galaxies with $\tau \le 2.5$ (i.e., consistent with old stellar populations with little on-going star formation) and with photometric redshifts in the range $0.3 \le z < 0.9$ (see Brown et al. in prep. for further details). Figure~\ref{fig:photoz} shows that the photometric redshifts in this range are relatively robust: of the 3056 red galaxies with spectroscopic redshifts, 2726 (89\%) have a difference of less than 0.1 between their photometric and spectroscopic redshifts.

 
\section{Analysis and Results}

\begin{figure*}
\begin{center}
\setlength{\unitlength}{1mm}
\begin{picture}(150,60)
\put(-15,-10){\includegraphics{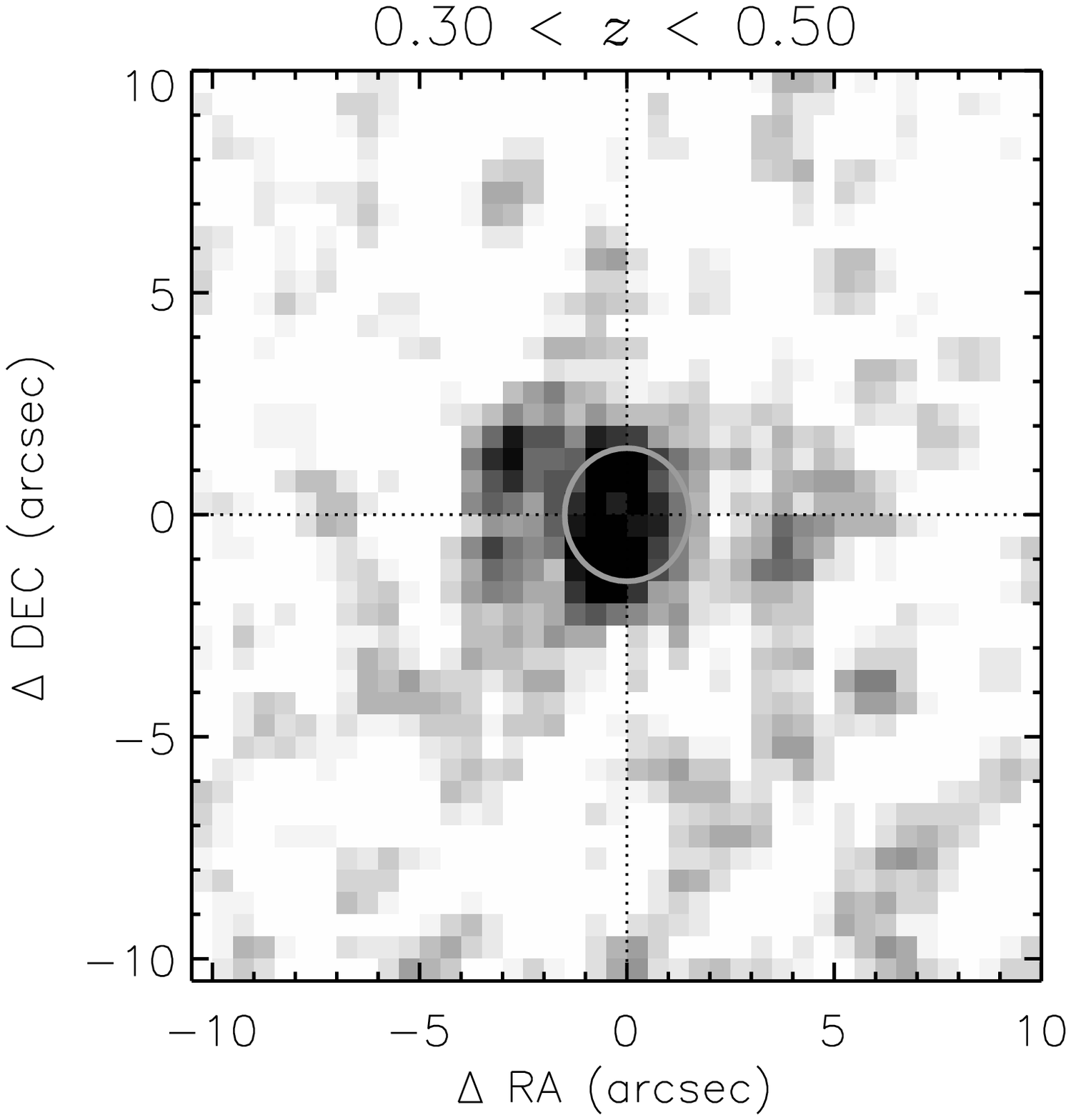}}
\put(40,-10){\includegraphics{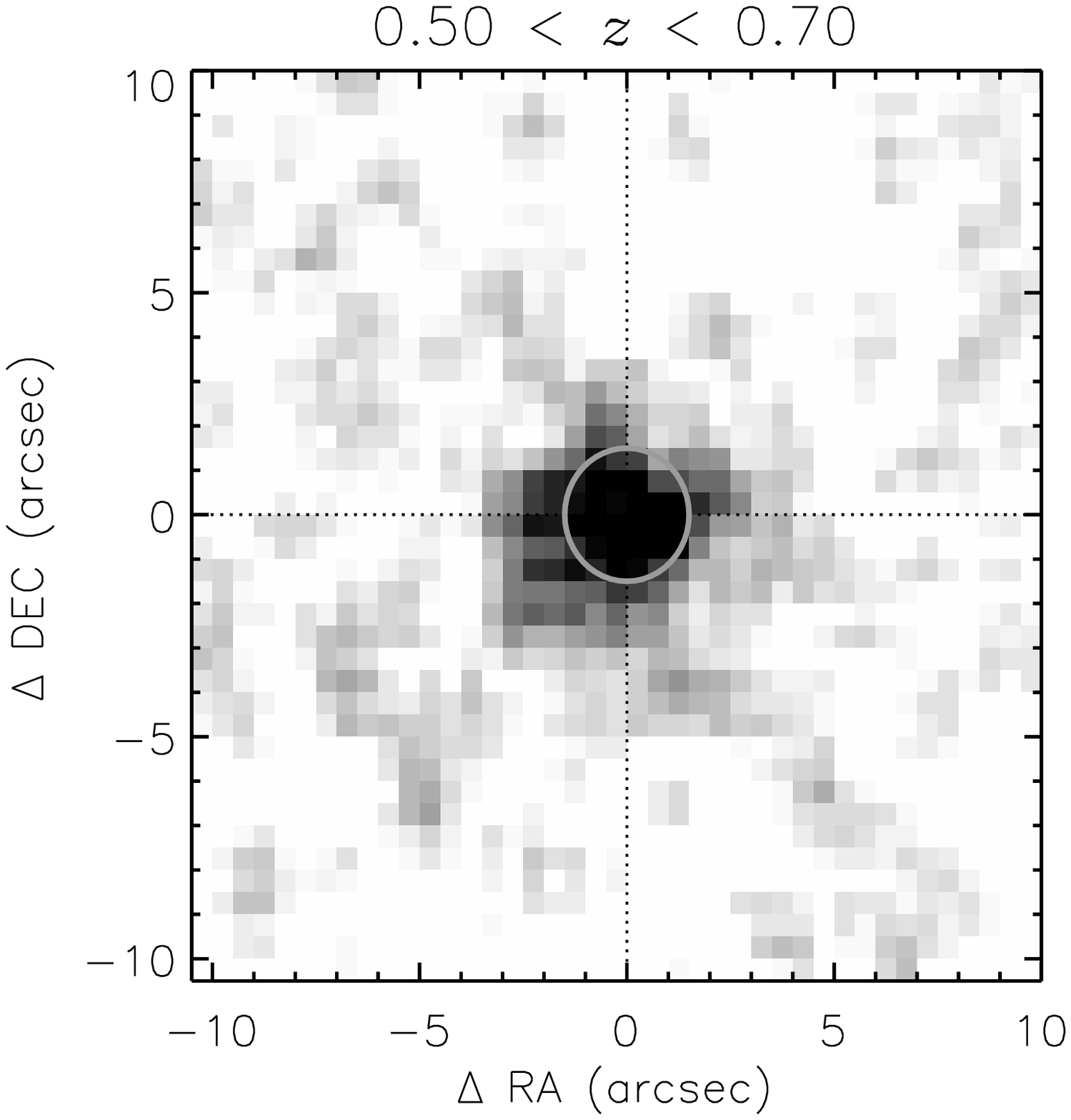}}
\put(95,-10){\includegraphics{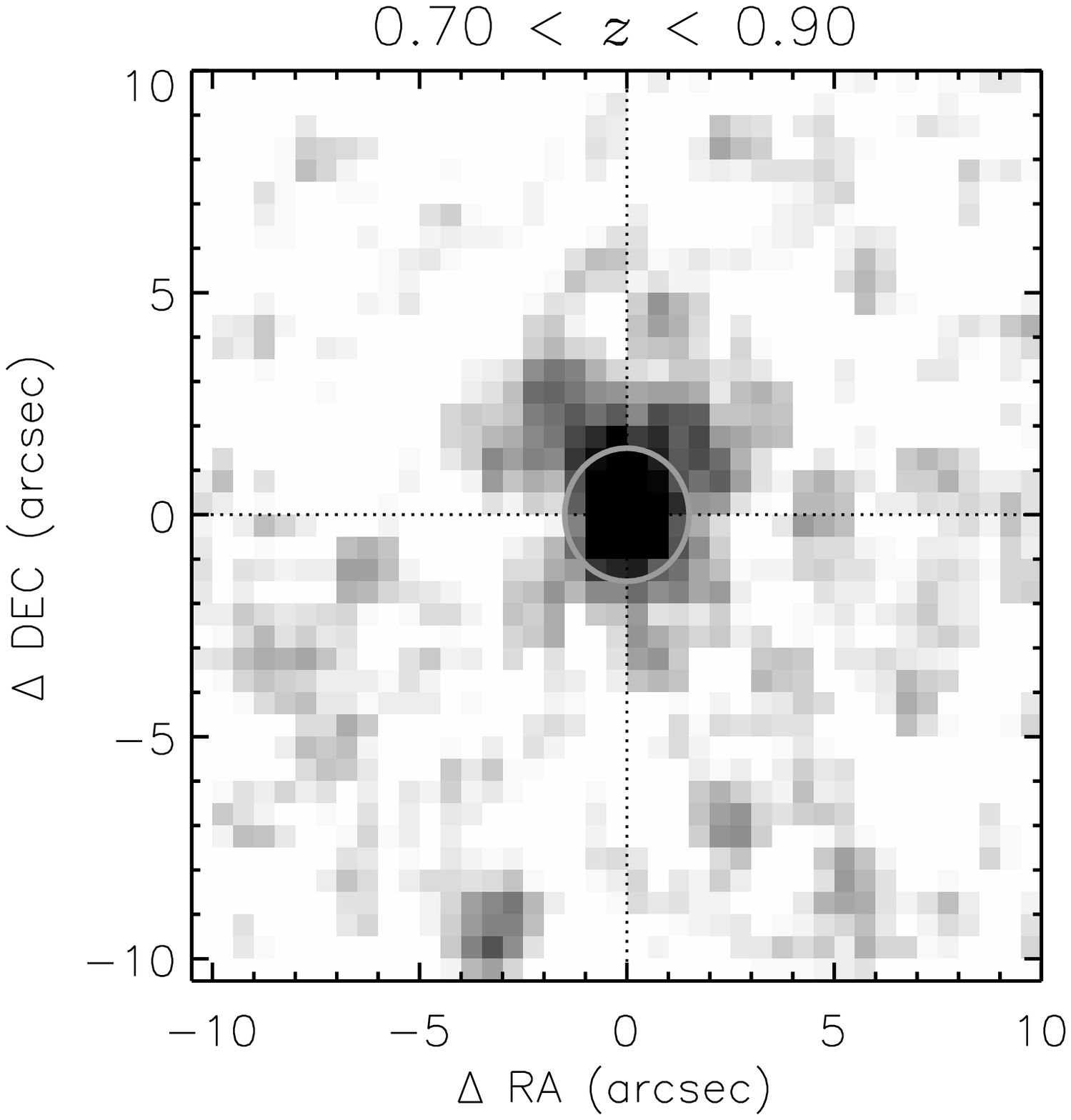}}
\end{picture}
\end{center}
{\caption[junk]{\label{fig:stack_im} The stacked full-band (0.5-7 keV) X-ray images of the individually undetected red galaxies with photometric redshifts $0.3 \le z<0.5$ (left), $0.5 \le z<0.7$ (center), and $0.7 \le z<0.9$ (right). To ensure that the signal is not dominated by a small number of sources, we exclude sources which are individually detected in the X-ray. The images are smoothed by a box-car average of 3 pixels. Over-plotted is the 1.5 arcsec aperture radius used to extract the signal.}}
\end{figure*}

We followed the X-ray stacking technique of Brand et al.~(2005) to investigate the mean X-ray luminosity of red galaxies as a function of their redshift and their optical luminosity. By stacking the X-ray observations of galaxies, one can increase the effective observation depth on the mean object by a factor of the sample size: a sample of $10^4$ galaxies effectively increases the depth of our 5~ks $Chandra$ exposure to that of a 50~Msec observation on the mean object. We divide our sample into 3 redshift bins: $0.3\le z<0.5$, $0.5 \le z<0.7$, and $0.7 \le z<0.9$. Figure~\ref{fig:stack_im} shows that a highly significant X-ray signal is obtained in all 3 redshift bins even when all individually detected X-ray sources with 4 or more X-ray counts are excluded from analysis. The stacked X-ray luminosity from all our red galaxies with no individual X-ray detection is $\approx 2\times 10^{41}{\rm ergs~s^{-1}}$. 

The total X-ray emission from galaxies may contain contributions from a variety of sources: stellar objects, such as low-mass and high-mass X-ray binaries (hereafter LMXBs and HMXBs); diffuse hot gas;  and AGN. The LMXB population is long-lived and traces the old stellar population (and hence the stellar mass) in galaxies; in contrast, HMXBs have short lifetimes and the luminosity of this population reflects the star-formation rate in galaxies. We use the relations of Kim et al.~(2004) and Grimm, Gilfanov, \& Sunyaev (2003) to determine the expected X-ray contribution of LMXBs and HBXBs to the total X-ray luminosity for each red galaxy. The average combined X-ray luminosity from both LMXBs and HMXBs is expected to be $\approx 2\times 10^{40}{\rm ergs~s^{-1}}$.

We conclude that the stacked X-ray luminosity is a factor of 10 higher than that expected from stellar sources alone, suggesting that the AGN dominates the X-ray emission in these sources. We note that the mean X-ray luminosity is a factor of $\sim$100 times smaller than the X-ray luminosities of typical Seyfert galaxies. This implies that although the X-ray luminosity is dominated by AGN emission, the accretion rate and/or accretion efficiency must be very low.

The large mean hardness ratio of our sample of red galaxies suggests that the X-ray spectrum is hard in comparison to typical unobscured AGN and is best fit by either an absorbed ($N_H\sim 2 \times 10^{22} {\rm cm^{-2}}$) $\Gamma$=1.7 power-law or an unabsorbed $\Gamma$=0.7 power-law consistent with models in which the fueling of the SMBH occurs at low accretion efficiencies (e.g., ADAF models; Rees et al.~1982; Di Matteo et al.~2000). Because hot gas will contribute primarily to the soft X-ray emission and the mean red galaxy spectrum is hard, we assume no contribution from hot gas to the total emission of stellar sources. 

\subsection{The X-ray Luminosity as a function of Optical Luminosity and Stellar Mass}

\begin{figure*}
\begin{center}
\setlength{\unitlength}{1mm}
\begin{picture}(150,75)
\put(-20,-15){\includegraphics{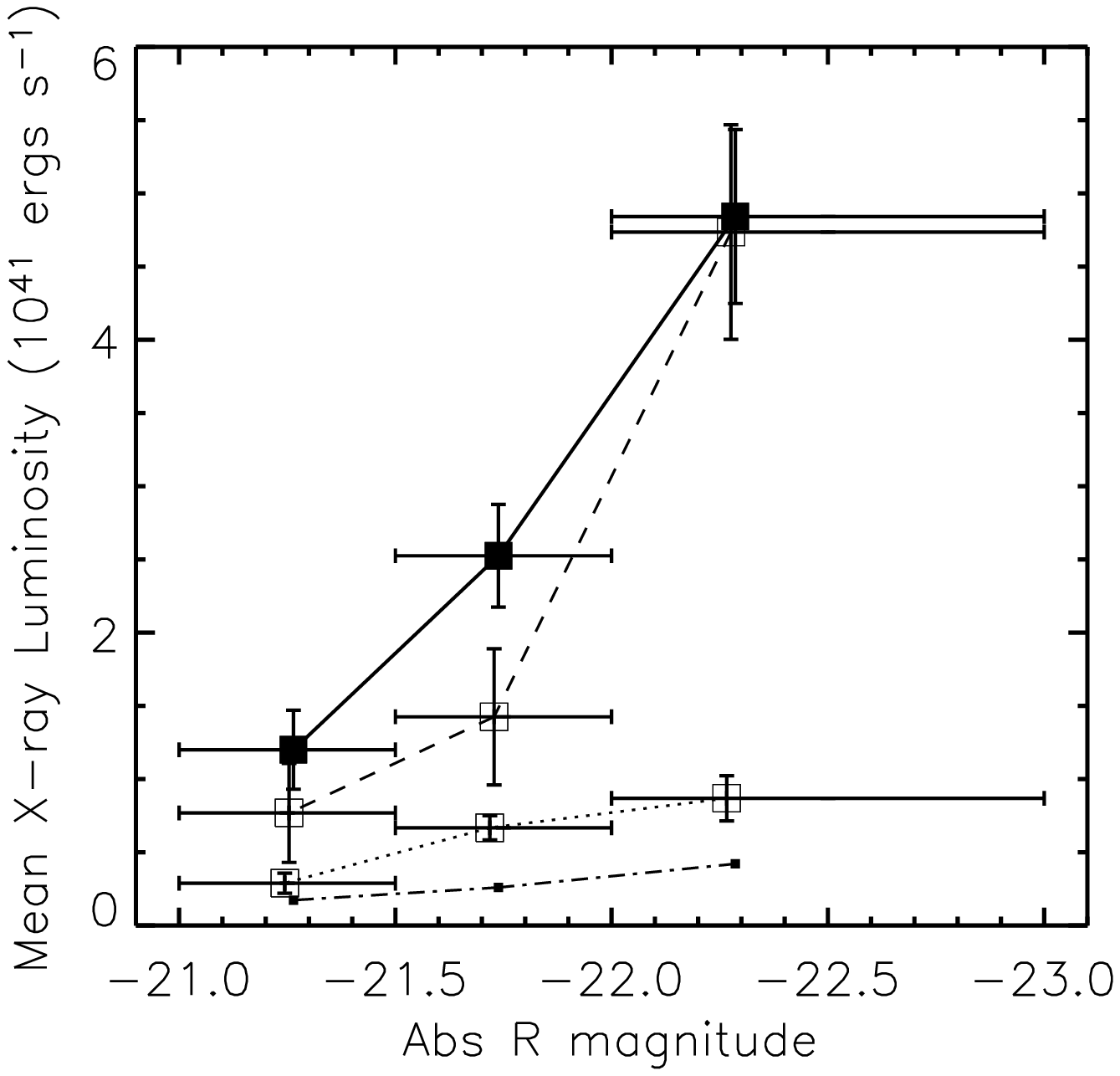}}
\put(60,-15){\includegraphics{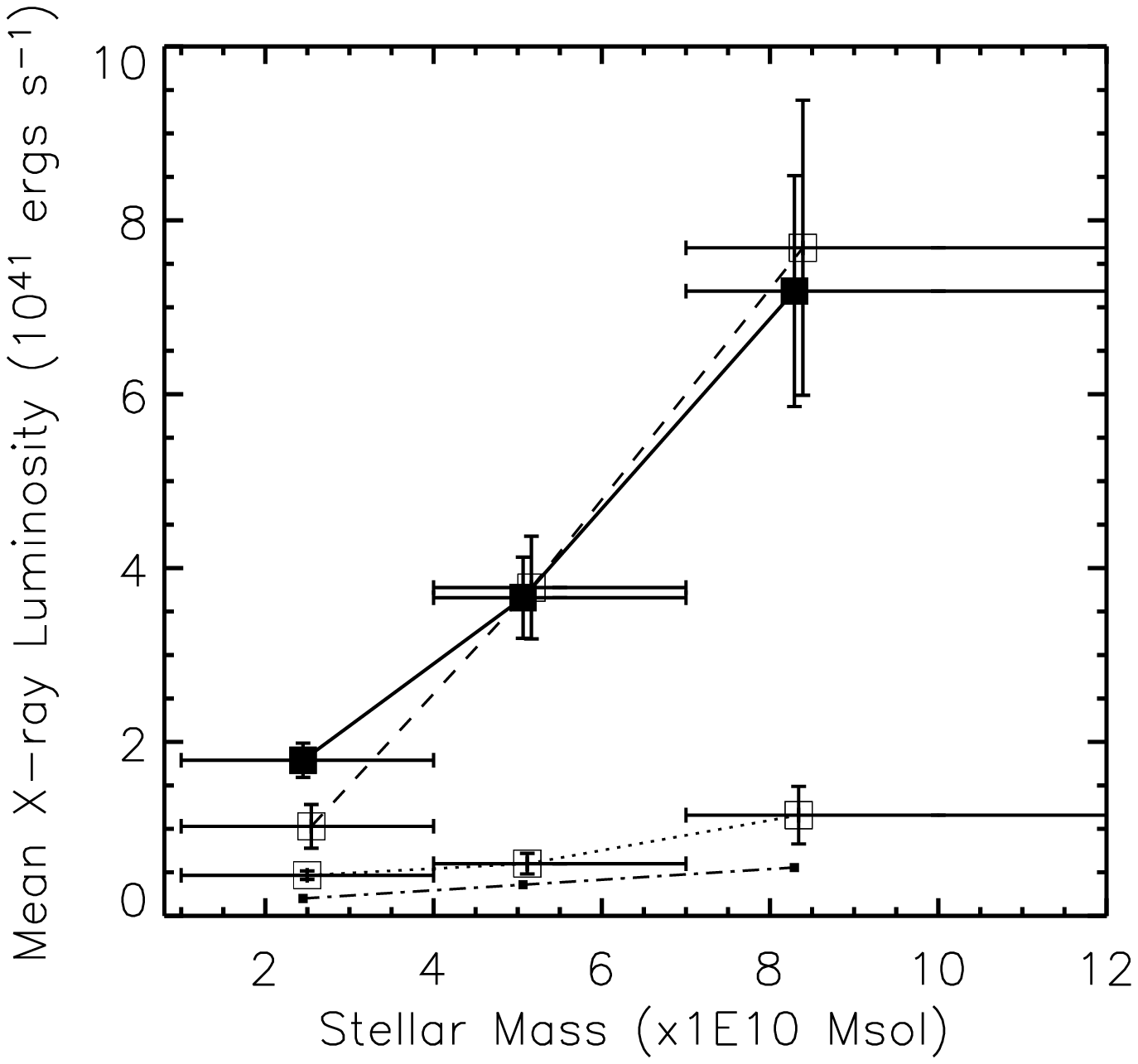}}
\put(10,60){\bf(a)}
\put(90,60){\bf(b)}
\end{picture}
\end{center}
{\caption[junk]{\label{fig:lxvsr} The mean X-ray luminosity and associated Poisson errors for the data split into 3 absolute $R$-band luminosity bins (a), and 3 stellar mass bins (b). The total, hard, and soft X-ray luminosities are represented by the solid, dashed, and dotted lines respectively. The X-ray flux is calculated assuming an absorbed ($N_H=2 \times 10^{22}~{\rm cm^{-2}}$) $\Gamma$=1.7 power-law spectral model. The hard band flux is probably slightly larger than the total-band flux in some cases because this model doesn't fit the true X-ray spectrum perfectly. To ensure that the signal is not dominated by a small number of sources, we exclude sources which are individually detected in the X-ray. The expected contribution from stellar sources is represented by the dot-dashed line.
}}
\end{figure*}

Figure~\ref{fig:lxvsr}a demonstrates that the mean X-ray luminosity increases as a function of absolute $R$-band magnitude (i.e. optically luminous galaxies contain more X-ray-luminous AGN). In figure~\ref{fig:lxvsr}b, we show that the mean X-ray luminosity also increases as a function of the stellar mass of the host galaxy. Interestingly, the hard-band X-ray emission appears to be responsible for the majority of the steep rise in the total-band X-ray luminosity at high stellar masses. Perhaps the most massive galaxies contain the largest reservoirs of gas which provide the fuel for a larger accretion rate but are also responsible for more absorption of the soft X-ray emission. Alternatively, the lower luminosity sources may have a higher contamination from stellar components which may boost the soft-band flux. 

\subsection{The Redshift Evolution of the Mean X-ray Luminosity}

\begin{figure*}
\begin{center}
\setlength{\unitlength}{1mm}
\begin{picture}(150,75)
\put(-25,-15){\includegraphics{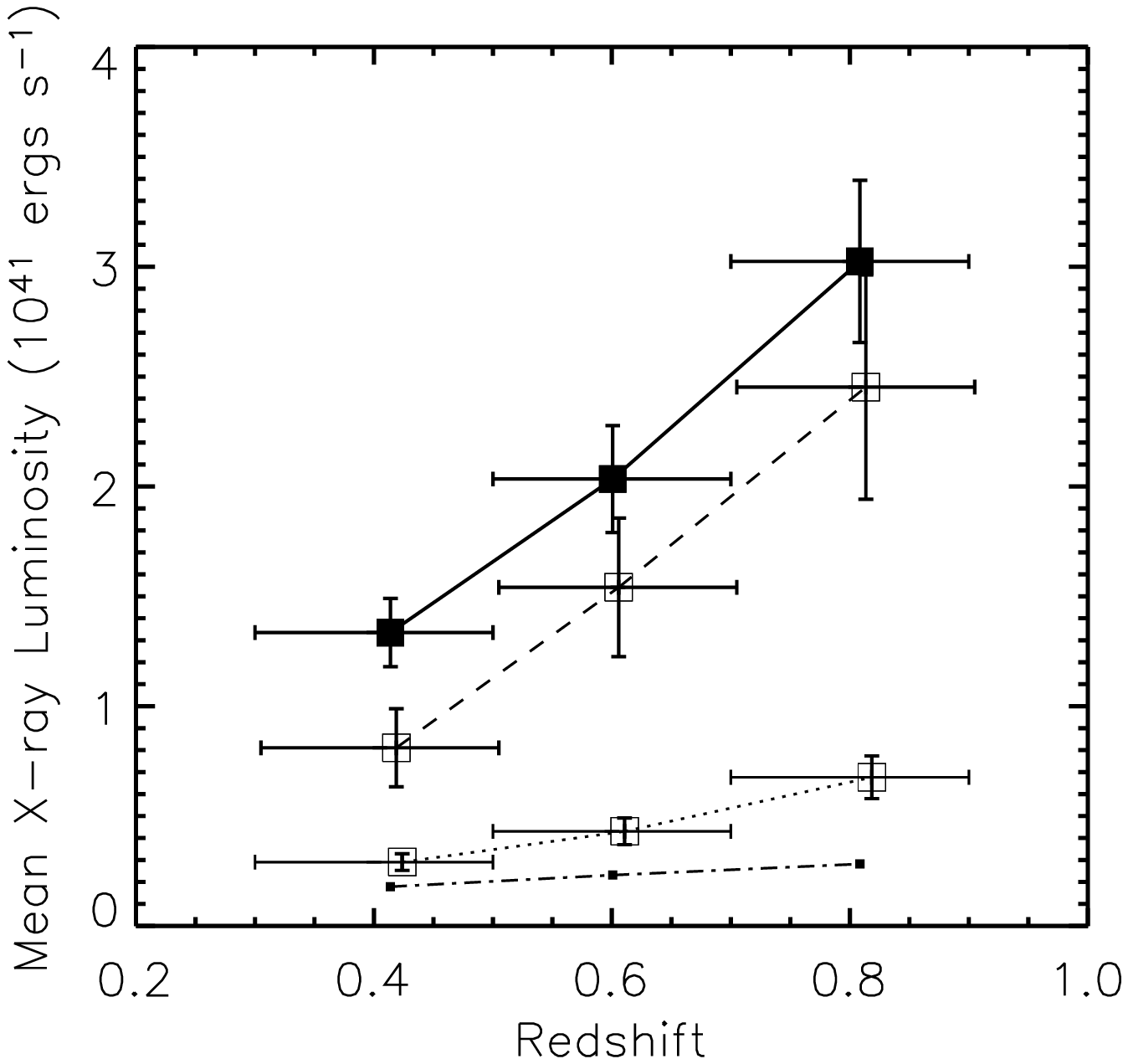}}
\put(65,-15){\includegraphics{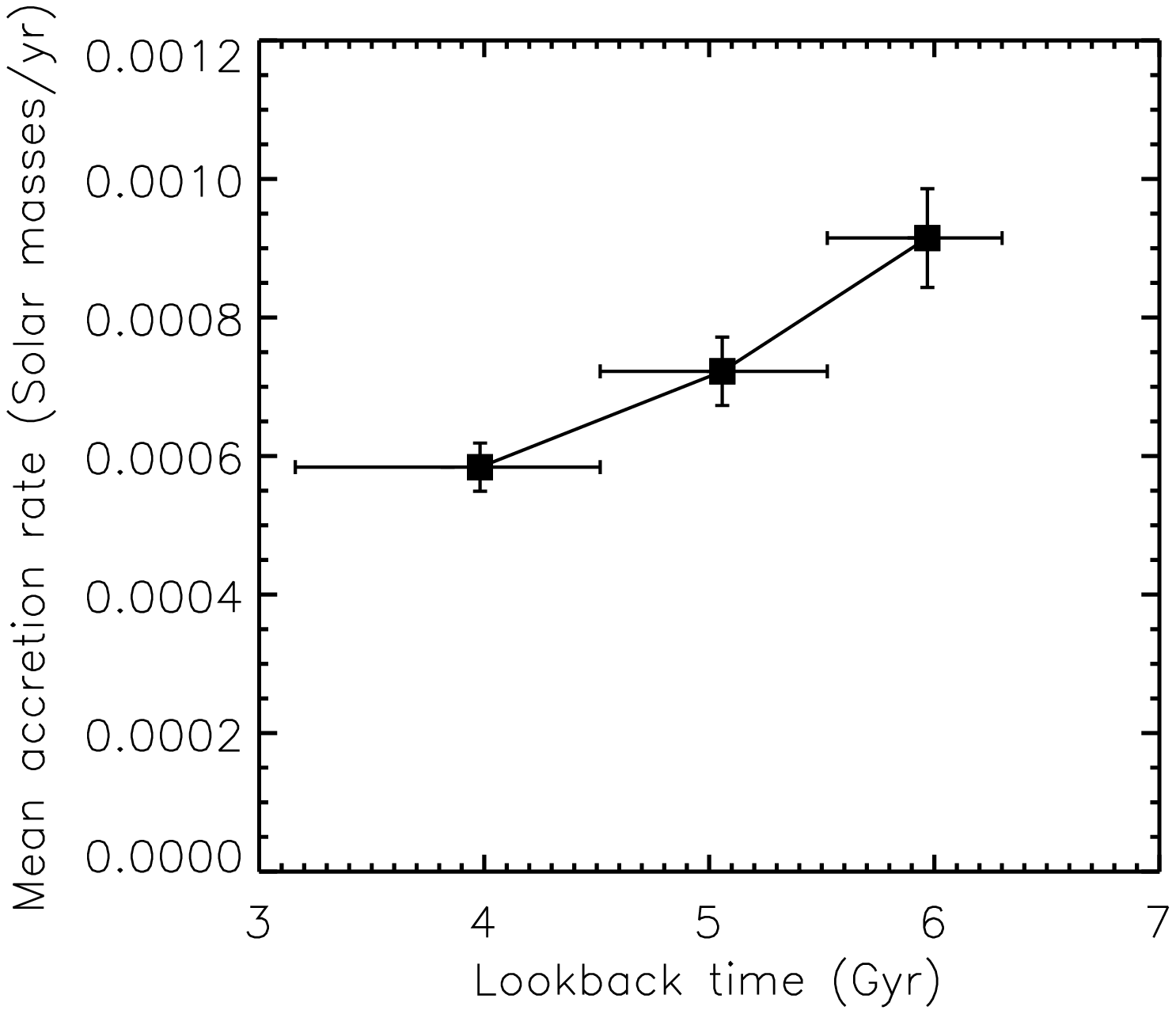}}
\put(5,60){\bf(a)}
\put(95,60){\bf(b)}
\end{picture}
\end{center}
{\caption[junk]{\label{fig:lxvsz} The mean X-ray luminosity as a function of redshift (a) and the mean SMBH accretion rate as a function of look-back time (b) for the red galaxies assuming an accretion efficiency of $\epsilon$=0.1.  The total, hard, and soft X-ray luminosities are represented by the solid, dashed, and dotted lines respectively. In calculating the mean accretion rate, we include the red galaxies whose X-ray emission is individually detected in the XBo\"otes survey. 
}}
\end{figure*}

Figure~\ref{fig:lxvsz}a shows the evolution of the mean X-ray luminosity of the red galaxy population as a function of redshift in the total, hard, and soft X-ray bands. The luminosities in all X-ray bands increase significantly with redshift and the results are not strongly dependent on whether we include the detected X-ray sources or not. We fit the relation to show that the total-band X-ray luminosity, $L_x\propto (1+z)^a$ where $a$=3.3$\pm$1.5. 

\subsection{The Mass Build-up of SMBHs}

If we assume that we know the efficiency at which energy is converted to radiation and that the X-ray luminosity dominates the bolometric luminosity of the AGN, we can use the scaling of Barger et al.~(2001) to determine the black hole accretion rate for a given X-ray luminosity. We can then integrate the mean accretion rate as a function of look-back time (Figure~\ref{fig:lxvsz}b) to determine the typical build-up of SMBH masses within red galaxies. We assume a typical accretion efficiency of $\epsilon$=0.1 and also include the X-ray detected sources in this calculation since these sources will be going through a more active phase of SMBH growth. We infer that the total growth expected for a SMBH within a red galaxy between $z\sim$1 and the present is $\approx 4 \times 10^5 ~{\rm M_{\odot}}$ (and $\approx 2 \times 10^5 ~{\rm M_{\odot}}$ when we only consider the sources that are not individually detected in the X-ray). The extrapolation of our data to $z=0$ appears justified given the results of Watson et al.~(2005) who infer a similar redshift evolution ($L_x\propto (1+z)^{3.4}$) for the mean X-ray luminosity in brighter ($I<20$) early type galaxies at $z=0.1-0.5$. The Magorrian relation (Magorrian et al.~1998) implies that our sample of red galaxies host SMBHs with masses in the range $10^7-10^8 {\rm M_\odot}$. We therefore conclude that SMBHs within red galaxies are accreting only an insignificant proportion of their mass between $z$=1 and the present if they are accreting at typical efficiencies; the main growth period must have occurred at higher redshifts.

There are a number of caveats that may affect our calculation. Firstly, the accretion efficiency could be substantially smaller ($\epsilon\approx$0.001) if the accretion rates are small enough for the flow to switch into a radially inefficient mode such as an advection-dominated accretion flow (ADAF; e.g. Rees et al.~1982). If this was the case, then a large and arguably implausible fraction of the SMBH mass could have been built up between $z\approx 1$ and the present. It is also possible that SMBH growth occurs primarily within very active (optically luminous) AGN phases, and that sources going through these phases temporarily fall out of our red galaxy classification (because their optical SEDs become dominated by AGN emission). In this case, we will also have under-estimated their SMBH mass build-up due to our assumption of a steady-state accretion.

\section{Summary} 

We stacked the X-ray emission from a large sample of red galaxies at $0.3 \le z<0.9$ and showed that the mean X-ray luminosity is an order of magnitude larger than expected from stellar sources alone and is likely dominated by low-luminosity AGN activity. Over the range of absolute $R$-band magnitudes sampled by our survey (in which the optical luminosity increases by a factor of $\sim 2.5$), the mean X-ray luminosity increases by a factor of $\sim$5. We find a similar relationship between the X-ray luminosity and stellar mass, implying that more massive galaxies have higher nuclear accretion rates or radiation efficiency than lower mass galaxies. We fit our results and determine that the mean X-ray luminosity evolves as $(1+z)^{3.3\pm1.5}$. We interpret this as a global decline in the mean AGN activity of normal early-type galaxies from $z\sim 1$ to the present in step with a similar decline in the global SFR. We integrate our fitted function to determine that if they are radiating with efficiencies typical of more luminous AGN, the SMBHs within red galaxies did not accrete a significant proportion of their mass in this time. 

\acknowledgements{Our research is supported by the National Optical Astronomy Observatory which is operated by the Association of Universities for Research in Astronomy, Inc. (AURA) under a cooperative agreement with the National Science Foundation. Support for this work was provided by the National Aeronautics and Space Administration through Chandra Award Number GO3-4176 issued by the $Chandra$ X-ray Observatory Center, which is operated by the Smithsonian Astrophysical Observatory for and on behalf of the National Aeronautics and Space Administration under contract NAS8-39073.
}

\vfill 
\end{document}